\documentclass[preprint,showpacs,preprintnumbers,amsmath,amssymb,nofootinbib]{revtex4-1}

\usepackage{graphicx}
\usepackage{dcolumn}
\usepackage{amssymb}
\usepackage{mathrsfs}
\usepackage{amsmath}
\usepackage{epsfig}
\usepackage{hhline}
\usepackage[utf8]{inputenc}
\usepackage{color}
\usepackage{multirow}
\usepackage[T1]{fontenc}
\usepackage{verbatim}
\usepackage{slashed}
\usepackage{epstopdf}
\usepackage{lineno,hyperref}
\usepackage{cancel}
\usepackage{xcolor}
\usepackage{fancyhdr}

\begin{document}

\title{Mono-top signature from stop decay at the HE-LHC}

\author{Xu-Xu Yang}
\author{Hang Zhou}
\email{Corresponding author: zhouhang@njnu.edu.cn}
\author{Tian-Peng Tang}
\author{Ning Liu}
\email{Corresponding author: liuning@njnu.edu.cn}

\affiliation{Department of Physics and Institute of Theoretical Physics, Nanjing Normal University, Nanjing, 210023, China}

\begin{abstract}
Searching for the top squark (stop) is a key task to test the naturalness of SUSY. Different from stop pair production, single stop production relies on its electroweak properties and can provide some unique signatures. Following the single production process $pp \to \tilde t_1 \tilde{\chi}^-_1 \to t \tilde{\chi}^0_1 \tilde{\chi}^-_1$, the top quark has two decay channels: leptonic channel and hadronic channel. In this paper, we probe the observability of these two channels in a simplified MSSM scenario. We find that, at the 27 TeV LHC with the integrated luminosity of ${\cal L} = 15~\text{ab}^{-1}$, $m_{\tilde{t}_1}<1900$ GeV and $\mu<750$ GeV can be excluded at $2\sigma$ through the leptonic mono-top channel, while $m_{\tilde{t}_1}<1200$ GeV and $\mu<350$ GeV can be excluded at $2\sigma$ through the hadronic channel.
\end{abstract}
\maketitle

\section{Introduction}
The discovery of Higgs~\cite{higgs-atlas,higgs-cms} made 2012 a landmark year of particle physics. It marks a significant success for the Standard Model (SM) and the efforts of particle physicists for nearly half a century have also been rewarded. This is very inspiring but there are still several dark clouds floating in the sky that make particle physicists disturbed and yearning at the same time, which all imply the existence of new physics and promote physicists to propose new models beyond the SM. Naturalness problem is one of the dark clouds and the weak scale supersymmetry (SUSY) is one of the most promising solutions. SUSY constructs a symmetry between fermions and bosons by introducing superpartners of the SM particles and then cancel the Higgs boson mass quadratic divergence naturally. We can make a bold assumption that $\mu<200$ GeV and $m_{\tilde{t}_{1,2}}<1.5$ TeV within the framework of MSSM~\cite{Baer:2012up,Han:2013usa,Tang:2019nyp,Yang:2002vz,Xiao:2006gu}. So if we want to test SUSY naturalness, searching for light stops in collider is very important~\cite{nsusy-1,nsusy-2,nsusy-3,nsusy-4,nsusy-7,nsusy-9,nsusy-12,Backovic:2015rwa,dutta,nsusy-14,nsusy-15,Goncalves:2016tft,Goncalves:2016nil,Wu:2018xiz,Abdughani:2018wrw,Duan:2017zar,Han:2016xet}.

\begin{figure}[htbp]
	\small
	\centering
	\includegraphics[width=5cm,height=3cm]{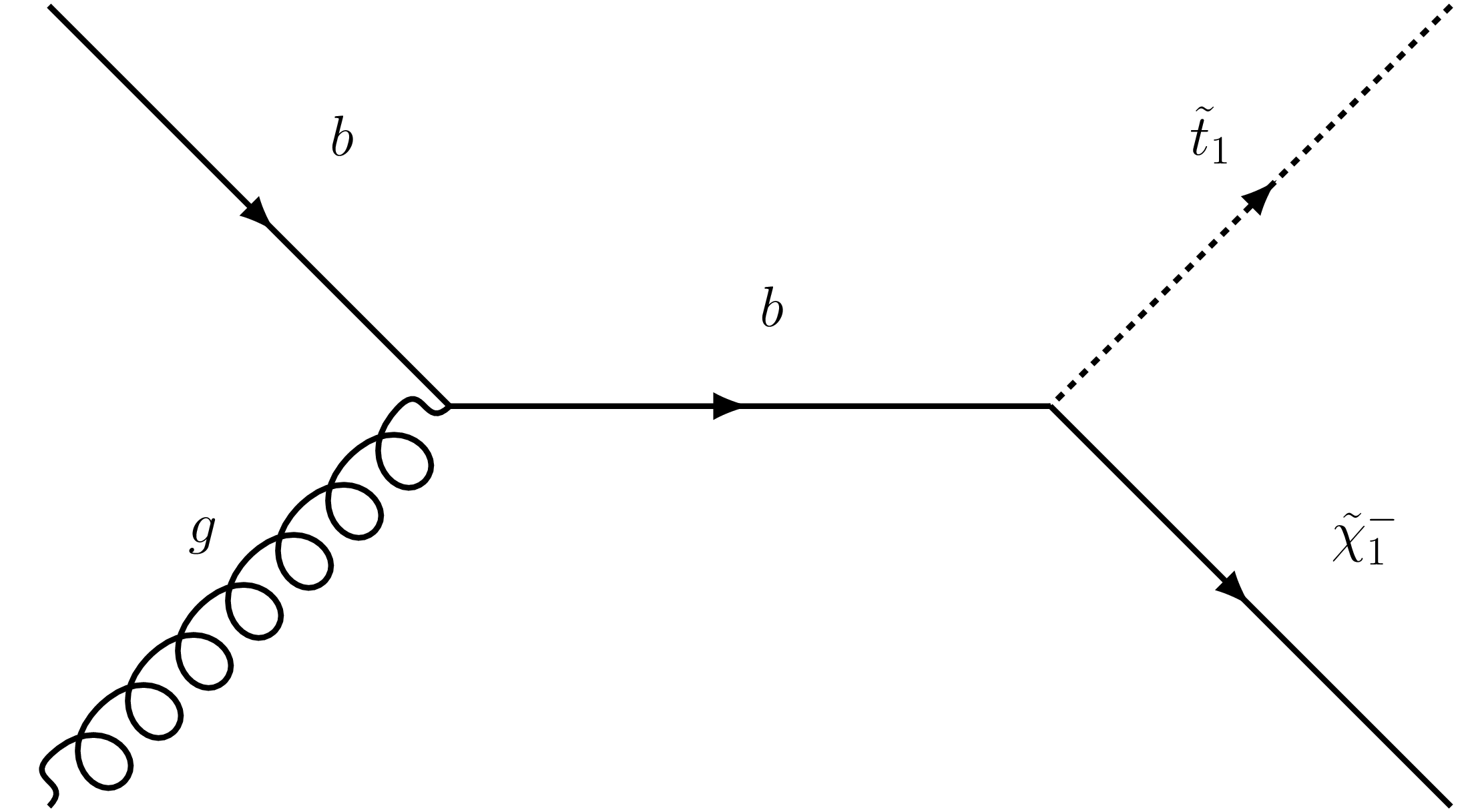}
	\includegraphics[width=5cm,height=3cm]{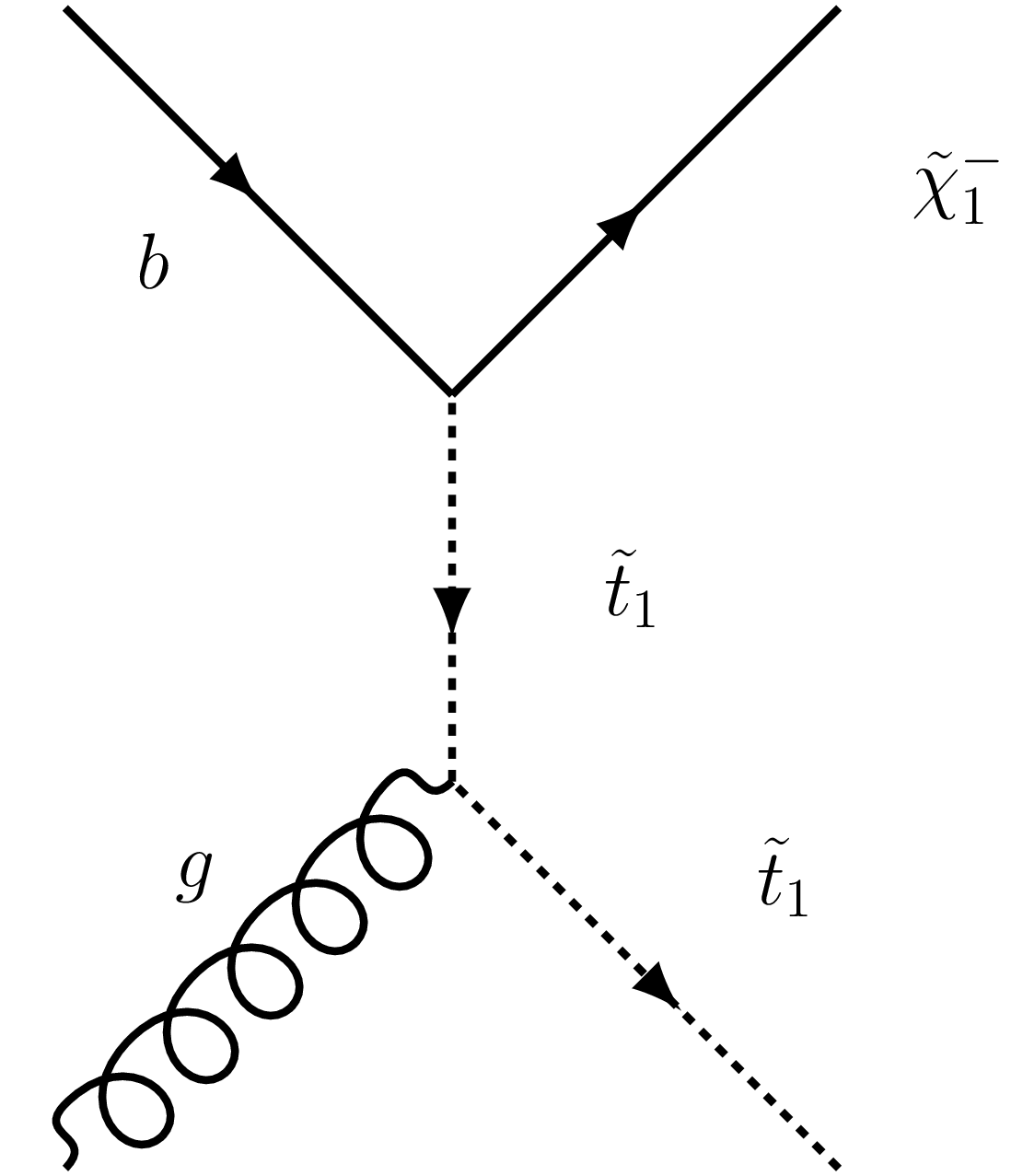}
	\caption{Feynman diagrams for the single stop production at LO.}
	\label{fig:feynman_diagram}
\end{figure}

LHC was built initially to search for Higgs boson but now we can also use it to search for new physics, including stops mentioned above. The null results of searching can exclude the low energy range and promote us to search for stops in a relatively high energy scale~\cite{Atlas,CMS}. Moreover, the large gap between the electroweak energy scale and the Planck energy scale drives the new physics to appear at TeV scale. Therefore, as the LHC cannot cater for the demand for new physics, proposals have been made to build High Energy LHC (HE-LHC), which is designed to run at collision energy of 27 TeV with the integrated luminosity of 15 ab$^{-1}$. In this work, we propose to search for stops through single stop production via electroweak interaction at the HE-LHC (\figurename~\ref{fig:feynman_diagram}). Following the production process, the stop can decay into the SM top quark which then can go through two decay channels: hadronic channel and leptonic channel. The search for stop by these two channels has been studied in Ref.~\cite{wu-1} at the 14 TeV LHC, we now explore their observability at the 27 TeV LHC. Generally, the stop pair production is considered as the discovery channel, but the single production can manifest the electroweak coupling between stop and neutralino, which can be used as a complementary study to the QCD pair production. After the stop mass can be determined through its pair production in the future, the single production search becomes a good way to identify the natures of electroweakinos. In addition, single production has distinct signatures at the LHC that are not present in pair production, which can serve as further confirmation of the discovery of stop~\cite{single-stop-1,single-stop-2,wu-1,wu-2}.

We assume a simplified scenario~\cite{snsusy-1,snsusy-2,snsusy-3,snsusy-4,snsusy-5} in our research that the only sparticles are the stop and charginos. Moreover, we set $\tan\beta = 50$ $m_{\tilde U_{3R}} \ll m_{\tilde Q_{3L}}$ and $\mu \ll M_{1,2}$, so that the top squark is right-handed and electroweakinos are higgsino-like. The branching ratio of $\tilde{t}_{R} \to t \tilde{\chi}^{0}_{1,2}$ is assumed to be 50\%~\cite{wu-2}. Because the mass splitting between $\tilde{\chi}^\pm_1$ and neutralino is very small, the leptons from $\tilde{\chi}^{\pm}_1$ decay are so soft that they will give a unique signature which is identified as transverse missing energy at the collider. We use the following softwares for Monte-Carlo simulation: \textsf{MadGraph5\_aMC@NLO} for event generation at parton level, \textsf{Pythia}~\cite{pythia} for hadronization and parton showering, \textsf{Delphes}~\cite{delphes} for detector simulation and \textsf{CheckMATE2}~\cite{checkmate} as a processor connecting the above tools. \textsf{SUSYHIT} is used to calculate the sparticles mass spectrum. To cluster jets, we use anti-$k_t$ algorithm with a cone radius $\Delta R = 0.4 $ ~\cite{antikt}.

This paper is organized as follows. In Sec.~II, we present the study of the leptonic channel from single stop production at the HE-LHC. And Sec.~III is on the hadronic channel. In Sec.~IV, we draw a conclusion on our work.

\section{Mono-top signature of leptonic channel from single stop production at the HE-LHC}
\begin{figure}[htbp]
	\centering
	\includegraphics[width=6.3cm,height=6.3cm]{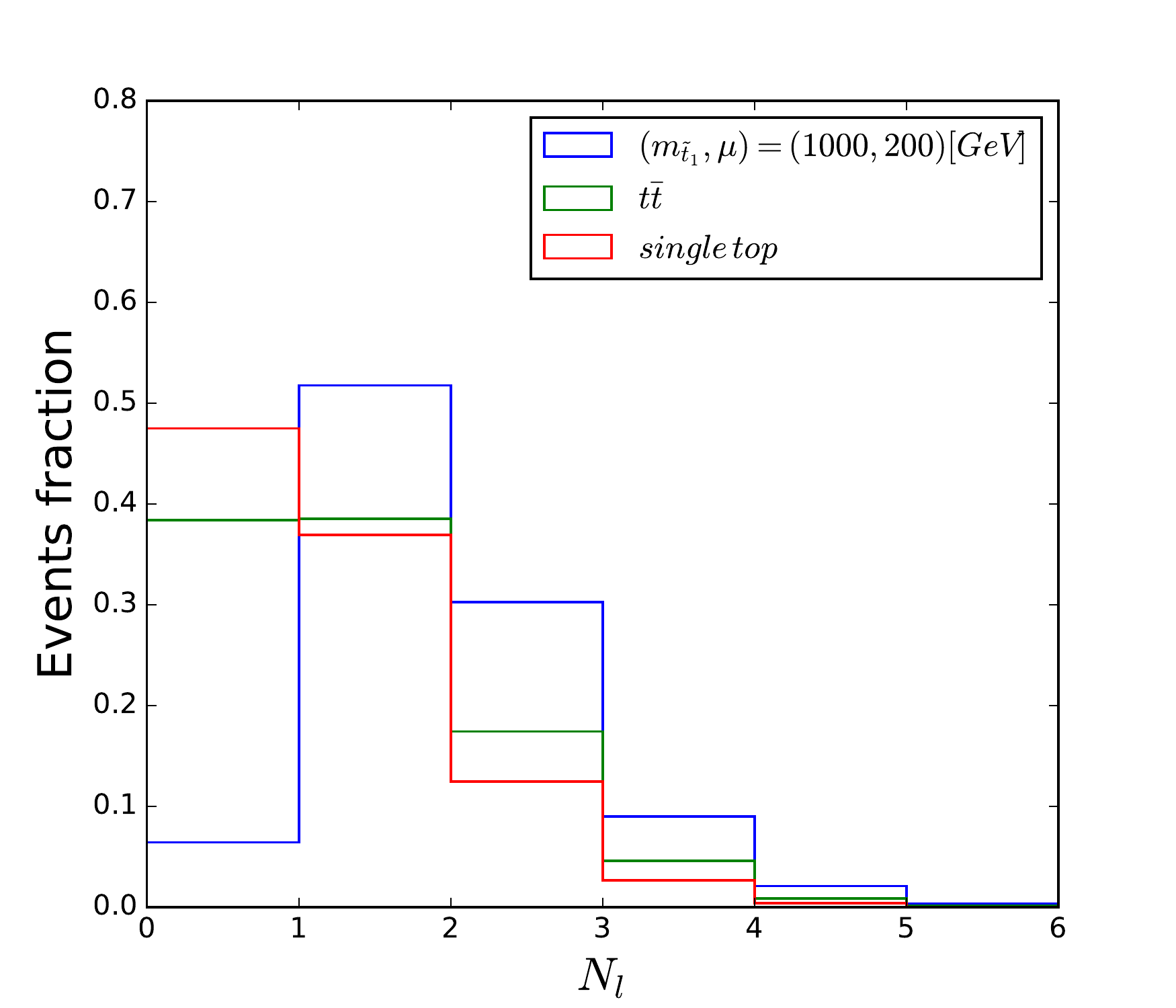}
	\includegraphics[width=6.3cm,height=6.3cm]{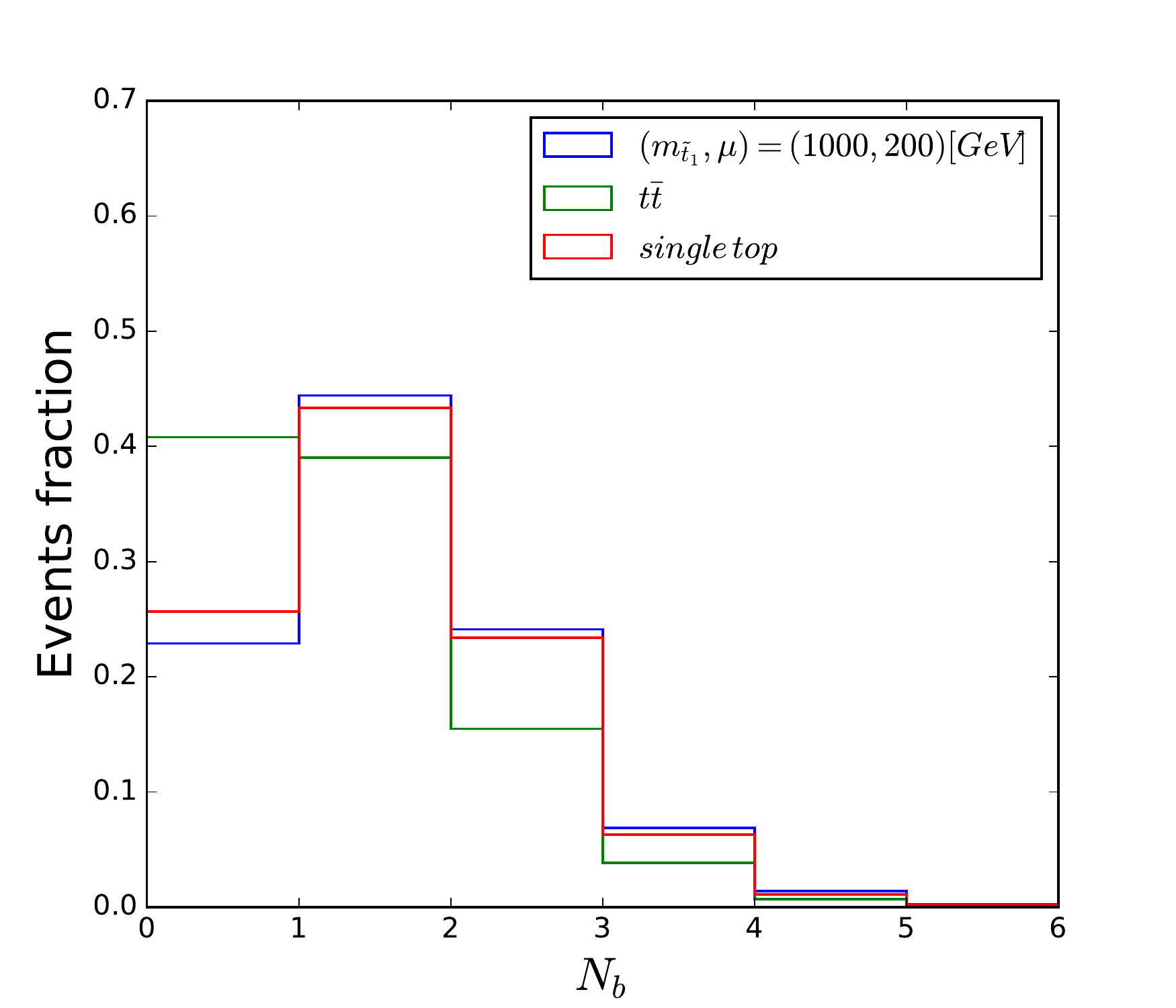}
	\includegraphics[width=6.3cm,height=6.3cm]{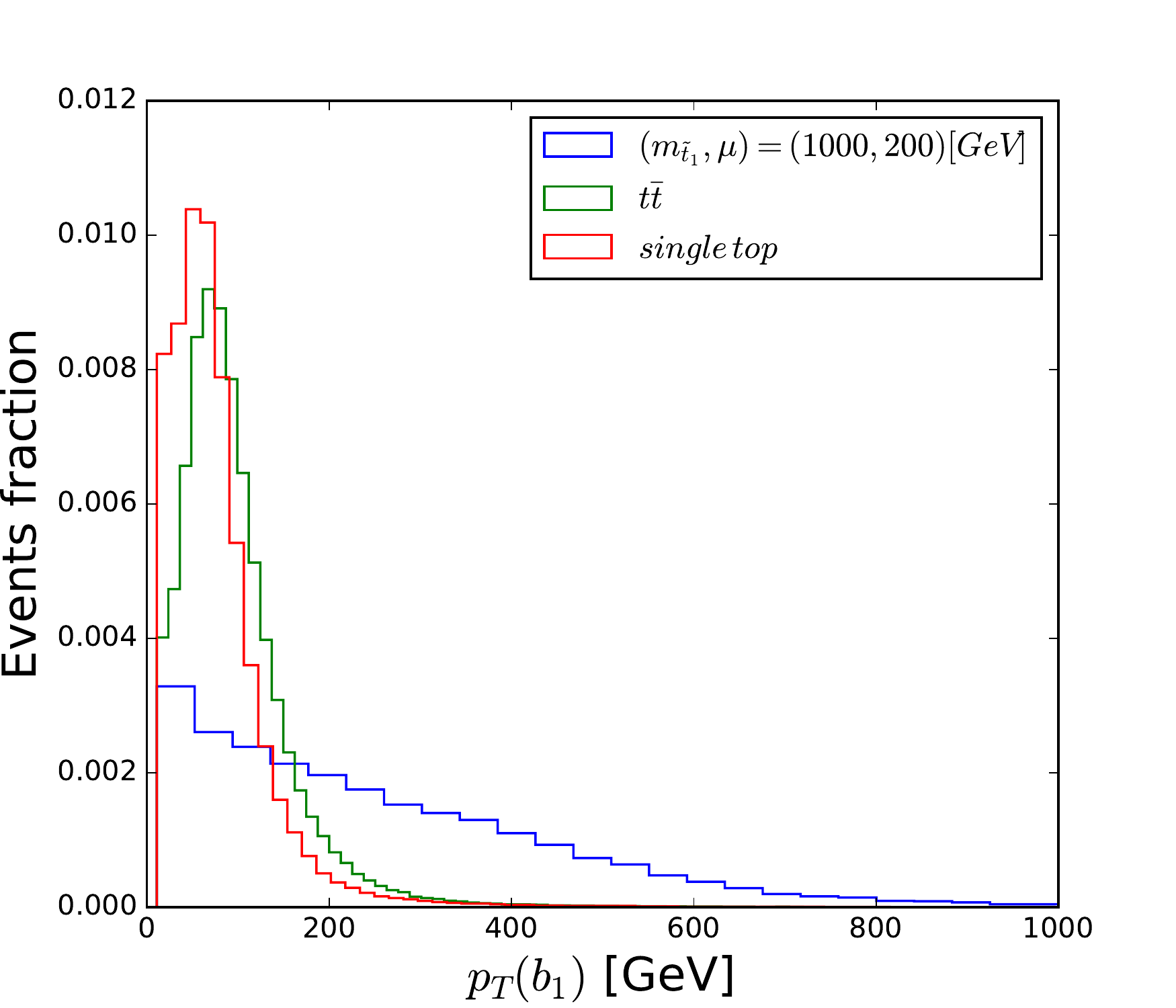}
	\includegraphics[width=6.3cm,height=6.3cm]{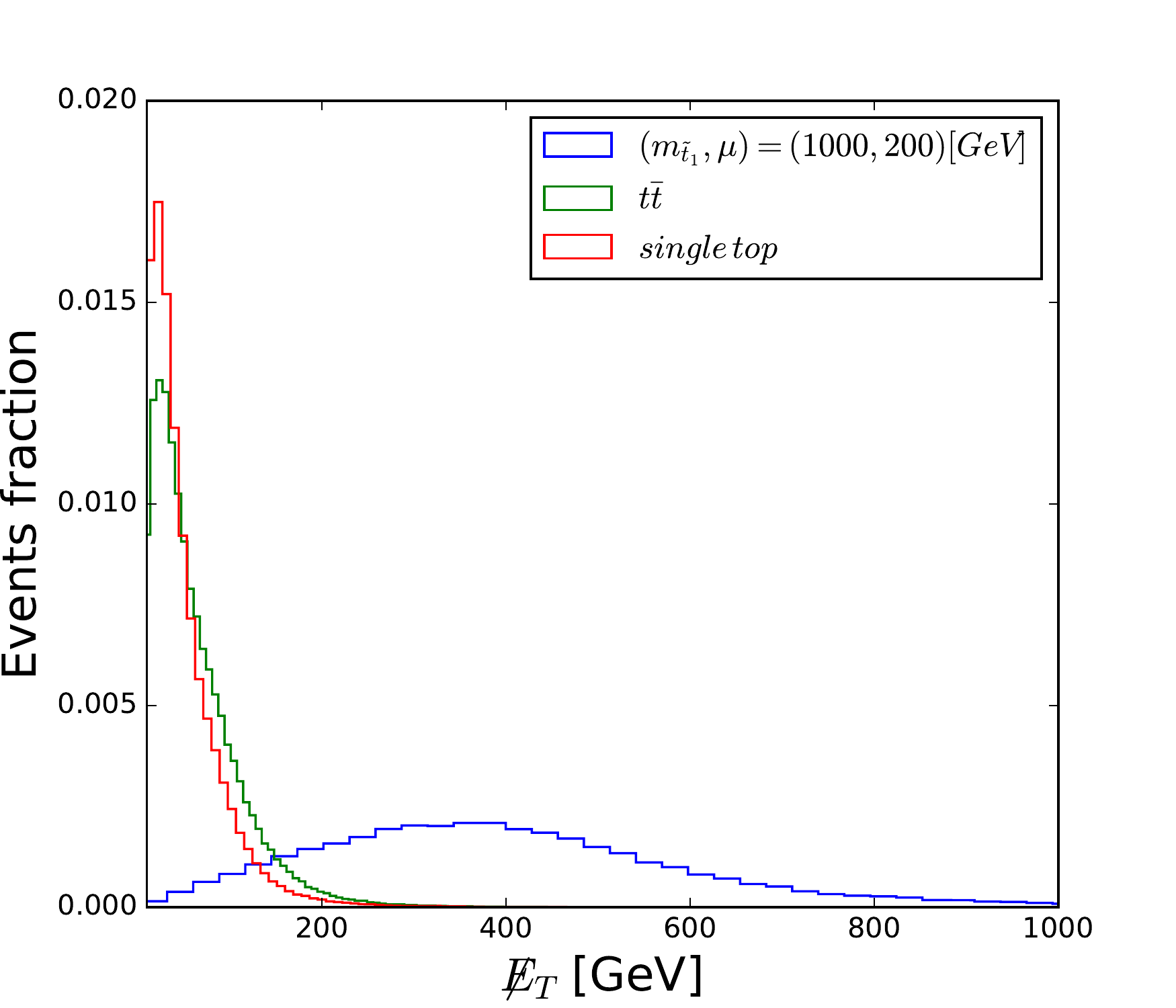}
	\includegraphics[width=6.3cm,height=6.3cm]{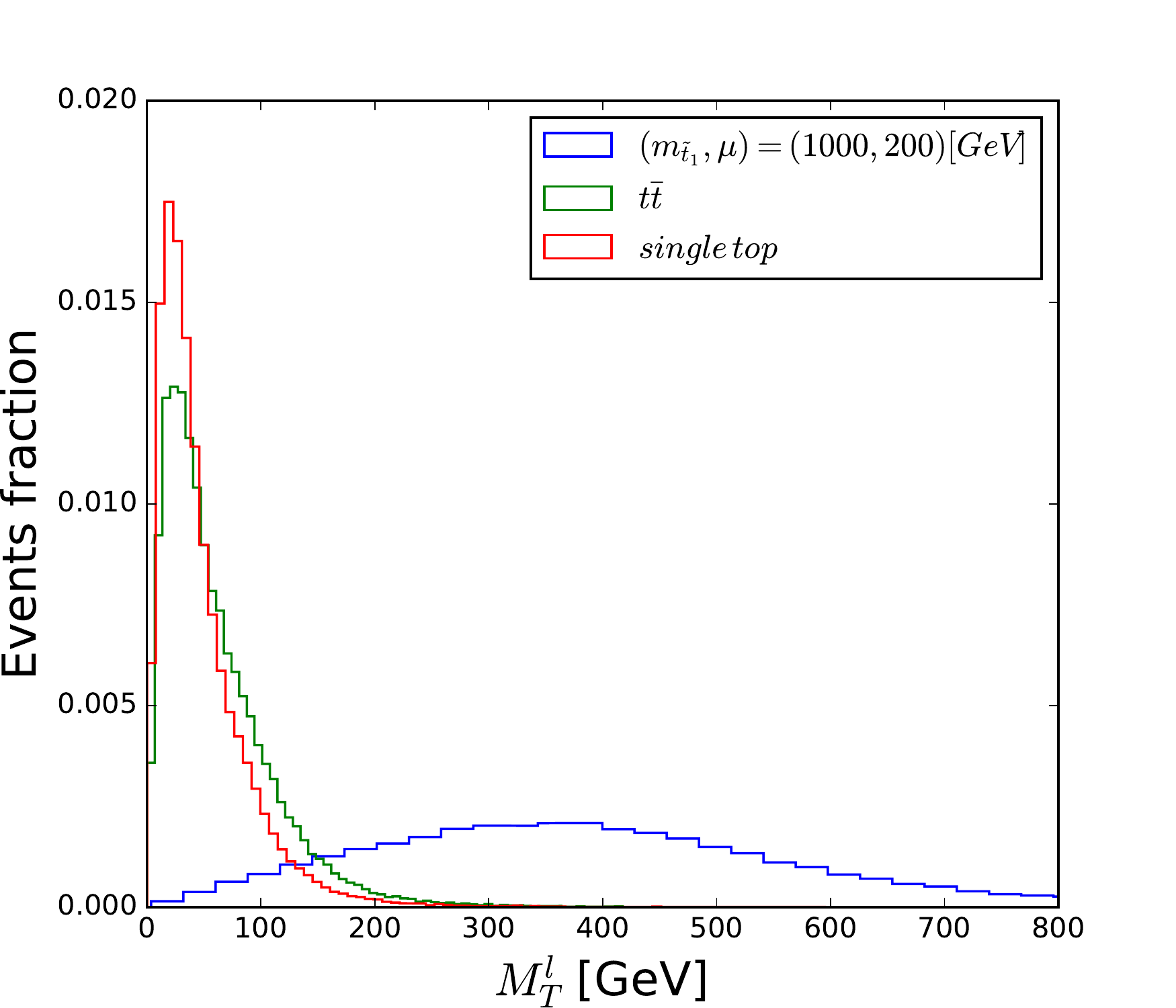}
	\includegraphics[width=6.3cm,height=6.3cm]{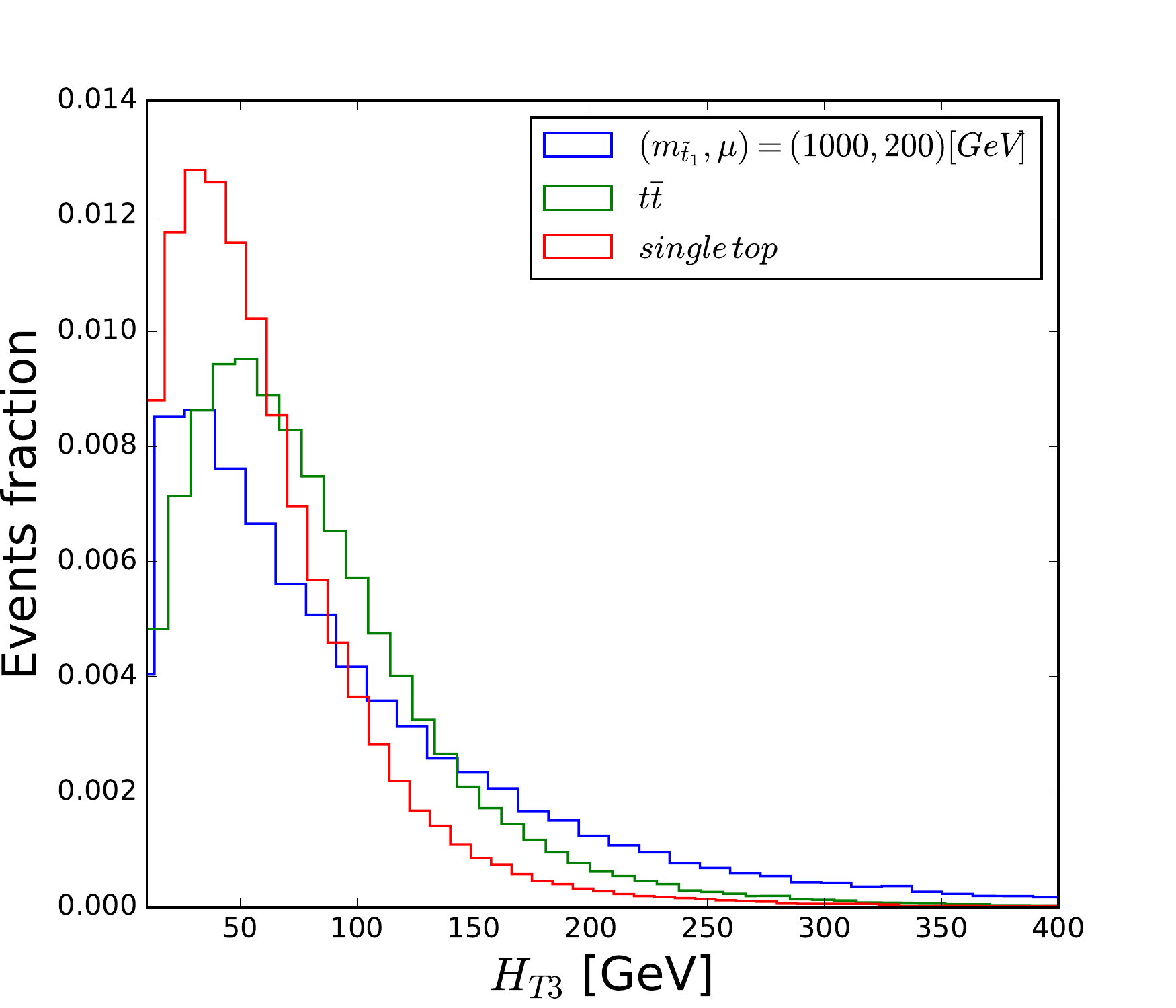}
	\caption{Normalized distributions of $N(l)$, $N(b)$, $p_{T}(b_{1})$, $\slashed E_T$, $M_T^l$, and $H_{T3}$ for the leptonic channel at 27 TeV LHC. The benchmark point is set as $\mu=200$ GeV and $m_{\tilde{t}_1}=1000$ GeV.}
	\label{fig:27TeV1}
\end{figure}
In this section, we investigate the mono-top signature of leptonic channel from single stop production:
\begin{eqnarray}
pp \to \tilde{t}_1 \tilde{\chi}^{-}_{1} \to t \tilde{\chi}^0_{1} \tilde{\chi}^-_1 \to \tilde{\chi}^0_{1} \tilde{\chi}^-_1 b l^+ \nu\,
\label{key}
\end{eqnarray}
The dominant background for this process is the SM $t\bar{t}$ production from semi- or full-leptonic decay of top quark, since the system of limited jet energy resolution and missing leptons can be miss-tagged as $\slashed E_T$. Another background comes from the single top production due to the undetected final leptons. We neglect some possible backgrounds, such as diboson production, because their cross sections are relatively small.
\begin{table}[t]
	\centering
	\begin{tabular}{|c|c|c|c|}
		\hline
		Cut & Signal & \multicolumn{2}{c|}{Background}\\
		\hline
		$m(\tilde{t}_{1}, \mu)$ [GeV] & (1000, 200) & single-top & $t\bar{t}$ \\
		\hline
		$N_{l} = 1$ & 8.179 & $3.1 \cdot 10^5$ & $8.75 \cdot 10^5$  \\
		\hline
		$N_{b} \ge 1$ & 6.296 & $2.31 \cdot 10^5$ & $8.08 \cdot 10^5$  \\
		\hline
		$p_{T}(b_{1}) > 200$ [GeV] & 3.262 & $7.9 \cdot 10^3$ & $4.45 \cdot 10^4$  \\
		\hline
		$\slashed{E}_T > 500$ [GeV] & 1.067 & 79.75 & 135.35  \\
		\hline
		$M_T^l > 650$ [GeV] & 0.473 & 37.89 & 39.63  \\
		\hline
		$H_{T3} < 200$ [GeV]       & 0.101 & 0.507 & 0.955  \\
		\hline
		$\Delta \phi(j,\slashed{\vec{p}}_T) > 0.6$ & 0.0524 & 0.2537 & 0.2387  \\
		\hline
	\end{tabular}
	\caption{A cut flow analysis of the cross sections (in unit of fb) for the leptonic channel events at the HE-LHC.}
	\label{tab:cutflow1}
\end{table}
The signature of Eq.~(\ref{key}) consists of a hard b jet and a charged lepton plus missing transverse energy $\slashed E_T$, from which we find several kinematic variables that can be used to separate signal and background, including the number of lepton $N_\ell$ and b-jet $N_b$, the transverse momentum of the leading b-jet $p_T(b_1)$, missing transverse energy $\slashed E_T$ and the transverse mass of lepton plus missing energy $M_T^l$. We also construct the variable $H_{T3}$~\cite{HT3} as the scalar sum of the transverse momentum of the third to fifth jet, to suppress the $t\bar{t}$ background, since the signal has less hard jets. We present the distributions of $N_\ell$, $N_b$, $p_T(b_1)$, $\slashed E_T$, $M_T^l$ and $H_{T3}$ for the signal and background events at the 27 TeV LHC. The benchmark point is $\mu=200$ GeV and $m_{\tilde{t}_1}=1000$ GeV. As shown in \figurename~\ref{fig:27TeV1}, lepton number of the signal events is more likely to center around $N_{\ell}=1$ while the backgrounds around $N_{\ell}=0$. Because of the boost effect of the b-jet from the stop decay, the leading b jets in signal events tend to be harder than that in the background events. The distributions of $\slashed E_T$ and $M_T^l$ in the larger region for the signal are natural consequences as a result of the massive $\tilde{\chi}^\pm_1$, which are well separated from the backgrounds.

After the above discussion, we select the events with the following cuts: (1) Exact one lepton in the final state is required; (2) We require at least one b-jet and that the leading $b$-jet satisfies $p_{T}(b_{1}) > 200$ GeV; (3) $\slashed E_T > 500$ GeV and $M_T^l > 650$ GeV are required; (4) $H_{T3} < 200$ GeV is required to further suppress the $t\bar{t}$ background; (5) To finally reduce multi-jet events in the $t\bar{t}$ background, we apply a cut as the minimum azimuthal angle between any of the jets and the missing transverse momentum $\Delta \phi(j,\slashed {\vec{p}}_T) > 0.6$.

We present the cutflow based on the above event cuts in Table~\ref{tab:cutflow1} for the leptonic channel of single stop production as well as for the SM backgrounds. As we can see from the table, the most effective cut is by the large $\slashed E_T$, which can suppress single-top and $t\bar{t}$ backgrounds by about ${\cal O} (10^2)$. To estimate the statistical significance $\alpha$, we use $\alpha=S/\sqrt{B}$ where S (B) is the events number after the above cuts applied. In FIG.~\ref{fig:contour12}\,(a), we present the $2\sigma$ exclusion limits for the leptonic channel on the plane of higgsino mass parameter $\mu$ versus stop mass $m_{\tilde{t}_1}$, from which we can find that $m_{\tilde{t}_1}<1900$ GeV and $\mu<750$ GeV can be excluded at $2\sigma$ level.
\begin{figure}[b]
	\centering
	\includegraphics[width=6.3cm,height=6.3cm]{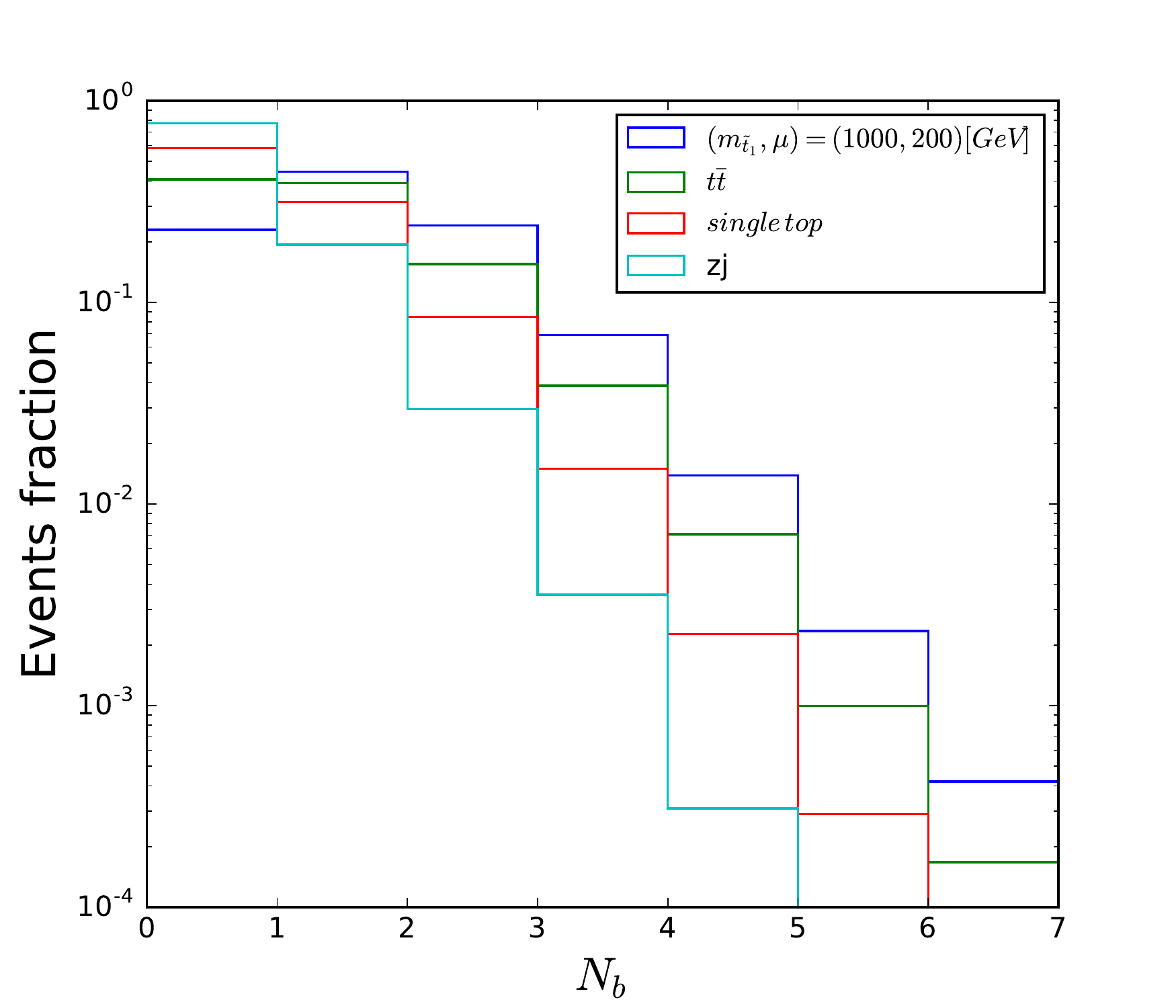}
	\includegraphics[width=6.3cm,height=6.3cm]{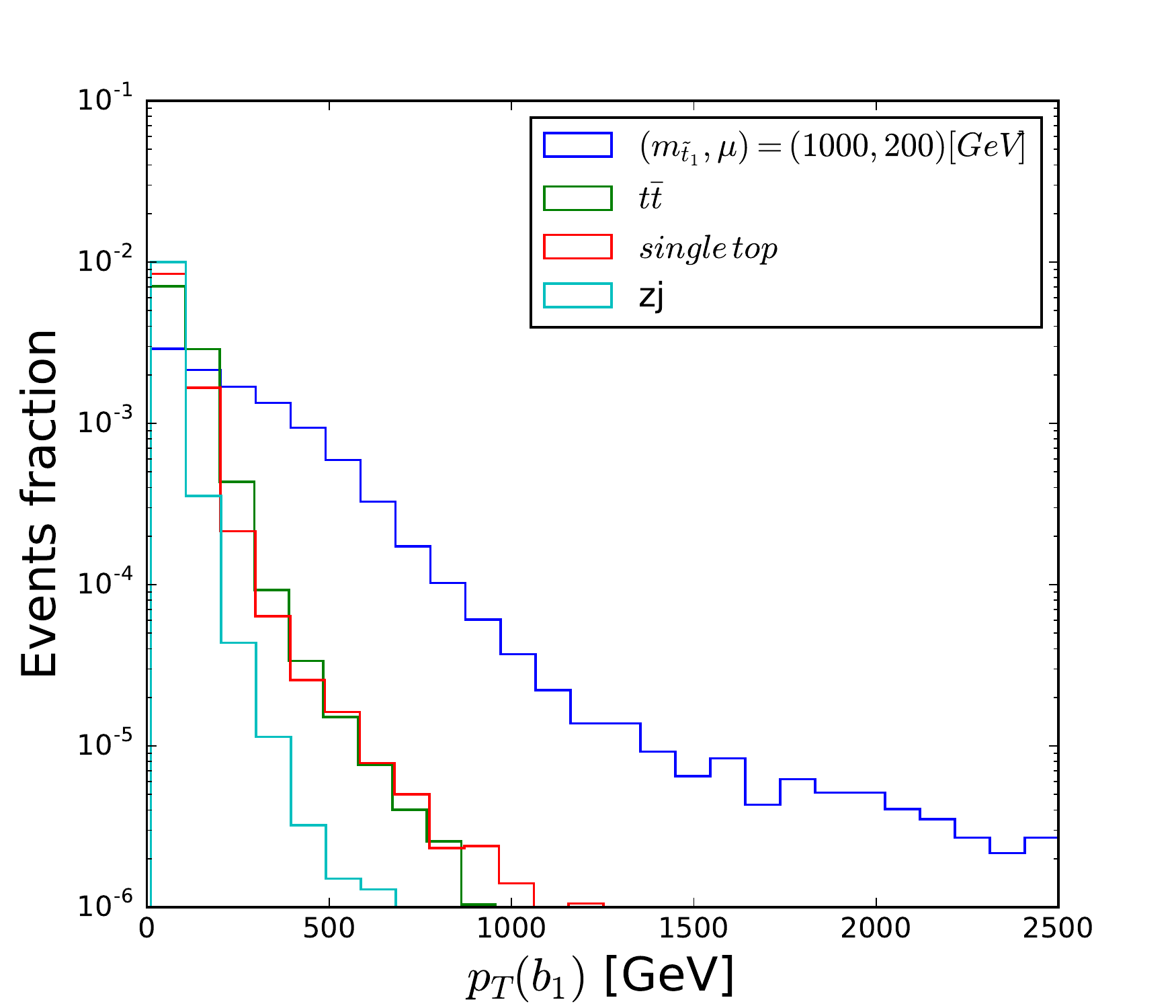}
	\includegraphics[width=6.3cm,height=6.3cm]{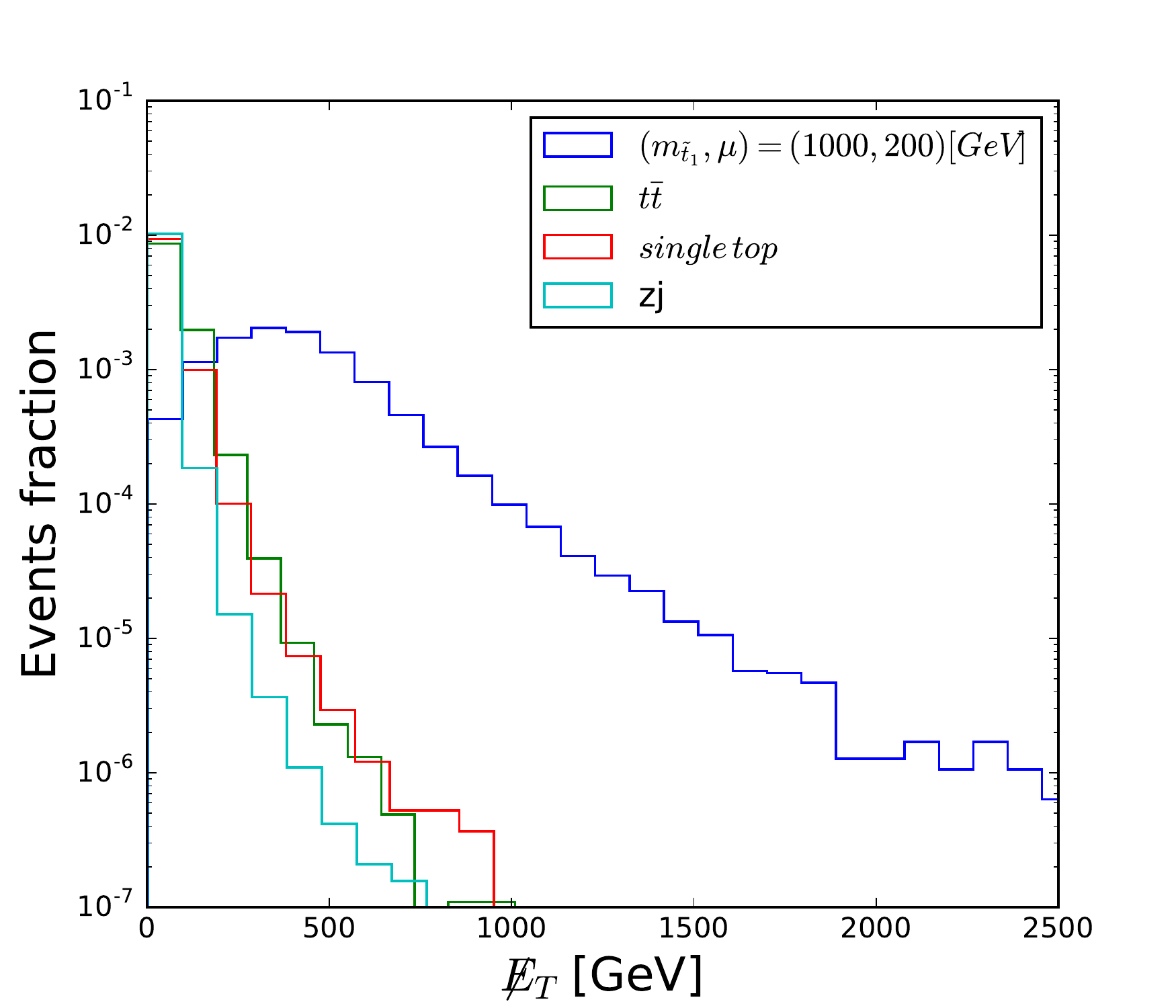}
	\includegraphics[width=6.3cm,height=6.3cm]{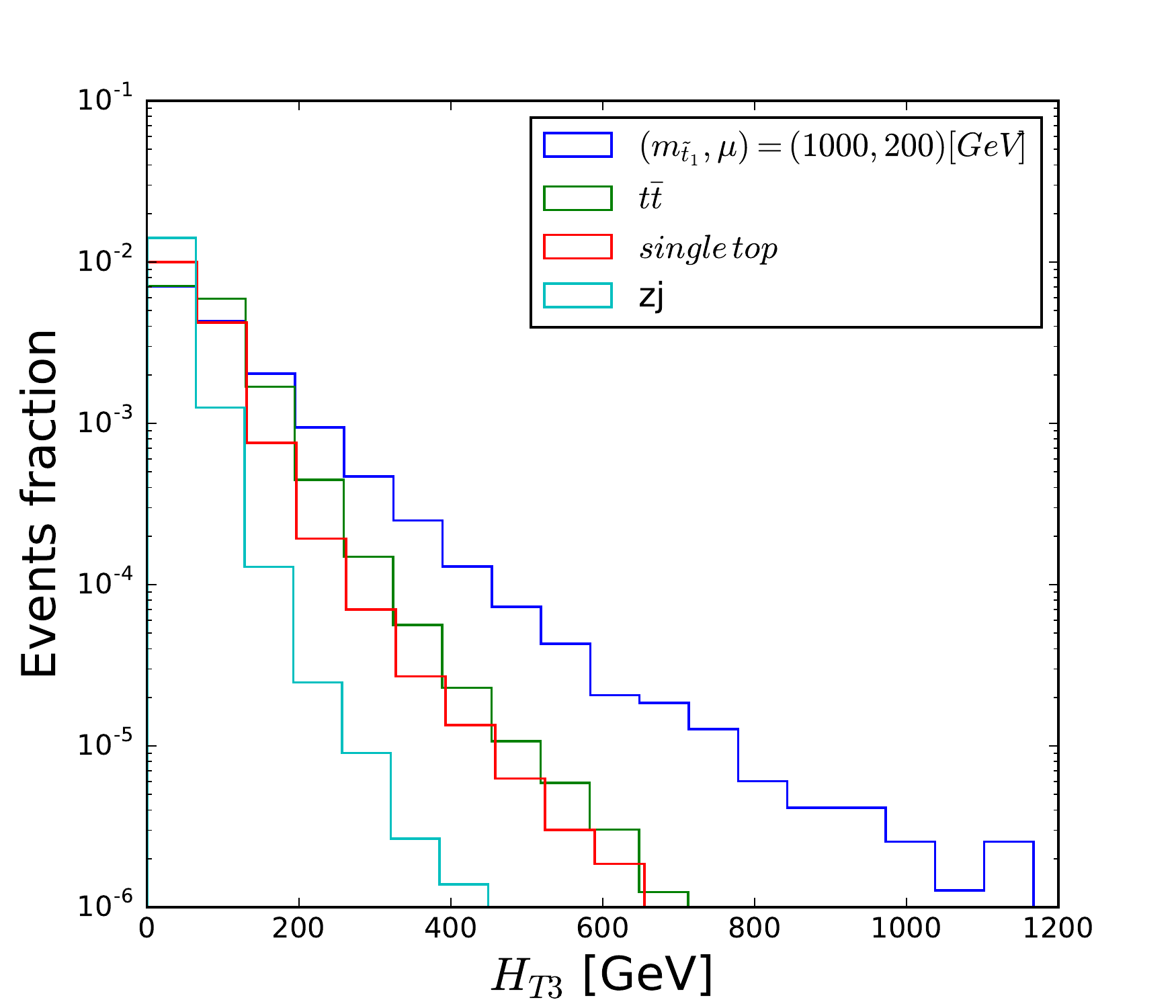}
	\caption{Normalized distributions of $N(b)$, $p_{T}(b_{1})$, $\slashed E_T$, and $H_{T3}$ for hadronic channel events at 27 TeV LHC. The benchmark point is set as $\mu=200$ GeV and $m_{\tilde{t}_1}=1000$ GeV.}
	\label{fig:distribution_14TeV2}
\end{figure}
\section{Mono-top signature of hadronic channel from single stop production at the HE-LHC}

In this section, we investigate the mono-top signature of hadronic channel from single stop production:
\begin{eqnarray}
pp \to \tilde{t}_1 \tilde{\chi}^{-}_{1} \to t \tilde{\chi}^0_{1} \tilde{\chi}^-_1 \to \tilde{\chi}^0_{1} \tilde{\chi}^-_1 b j j\,.
\end{eqnarray}
In addition to $t\bar{t}$ and single top background, another dominant background is $Z + {\mathrm jets}$ because light-flavor jets can fake the signal b-jet in the final state.

In \figurename~\ref{fig:distribution_14TeV2}, we present the distributions of $N_b$, $p_T(b_1)$, $\slashed E_T$ and $H_{T3}$ . From the kinematic distributions, we can find that signal has more events than background centered around $N_b = 1$. Similar to the case of leptonic channel, the boost of the leading b-jet from the stop decay leads to a larger $p_T(b_1)$ for the signal and the massive $\tilde{\chi}^\pm_1$ leads to a larger $\slashed E_T$ of the signal events than the backgrounds. $H_{T3}$ can also be used to distinguish between signal and background due to less hard jets in the signal.
\begin{table}[b]
	\centering
	\begin{tabular}{|c|c|c|c|c|}
		\hline
		Cut & Signal & \multicolumn{3}{c|}{Background}\\
		\hline
		$m(\tilde{t}_{1}, \mu)$ [GeV] & (1000, 200) & single-top & $t\bar{t}$ & $Z + {\mathrm jets}$\\
		\hline
		Lepton veto & 8.976 & $3.1 \cdot 10^5$ & $9.15 \cdot 10^5$ & $1.01 \cdot 10^8$\\
		\hline
		$N_{b} \ge 1$ & 6.863 & $2.34 \cdot 10^5$ & $7.65 \cdot 10^5$ & $2.15 \cdot 10^7$\\
		\hline
		$p_{T}(b_{1}) > 200$ [GeV] & 3.8 & $7.9 \cdot 10^3$ & $4.74 \cdot 10^4$ & $1.6 \cdot 10^5$\\
		\hline
		$\slashed{E}_T > 700$ [GeV] & 0.39 & 25.46 & 12.41 & 455  \\
		\hline
		$p_{T}(j_{1}) > 650$ [GeV] & 0.26 & 21.06 & 10.98 & 364  \\
		\hline
		$H_{T3} < 200$ [GeV]       & 0.085 & 1.35 & 1.19 & 156  \\
		\hline
		$\Delta \phi(j,\slashed{\vec{p}}_T) > 0.6$ & 0.045 & 0.93 & 1.19 & 65  \\
		\hline
	\end{tabular}
	\caption{A cut flow analysis of the cross sections (in unit of fb) for the hadronic channel events at the HE-LHC. Note that $S/B$ for this channel is tiny due to the large SM backgrounds and the sensitivity is worse than that of the leptonic channel, as shown in FIG.~\ref{fig:contour12}.}
	\label{tab:cutflow2}
\end{table}
Then we use the following cuts after the above analysis on the kinematic distributions: (1) We require no leptons in the final states; (2) We require at least one b-jet and the leading $b$-jet satisfies $p_{T}(b_{1}) > 200$ GeV; (3) $\slashed E_T > 700$ GeV is required; (4) We require $p_{T}(j_{1}) > 650$ GeV; (5) $H_{T3} < 350$ GeV is required; (6) $\Delta \phi(j,\slashed {\vec{p}}_T) > 0.6$ where $\Delta \phi(j,\slashed {\vec{p}}_T)$ is defined in the same way as in the last section.
In Table~\ref{tab:cutflow2}, we can see that the large $\slashed E_T$ requirement can suppress single-top and $t\bar{t}$ backgrounds most effectively by about ${\cal O} (10^3)$, with about 10\% signal events surviving.

\begin{figure}[t]
	\centering
	\begin{minipage}{0.48\linewidth}
		\centerline{\includegraphics[width=8cm,height=8cm]{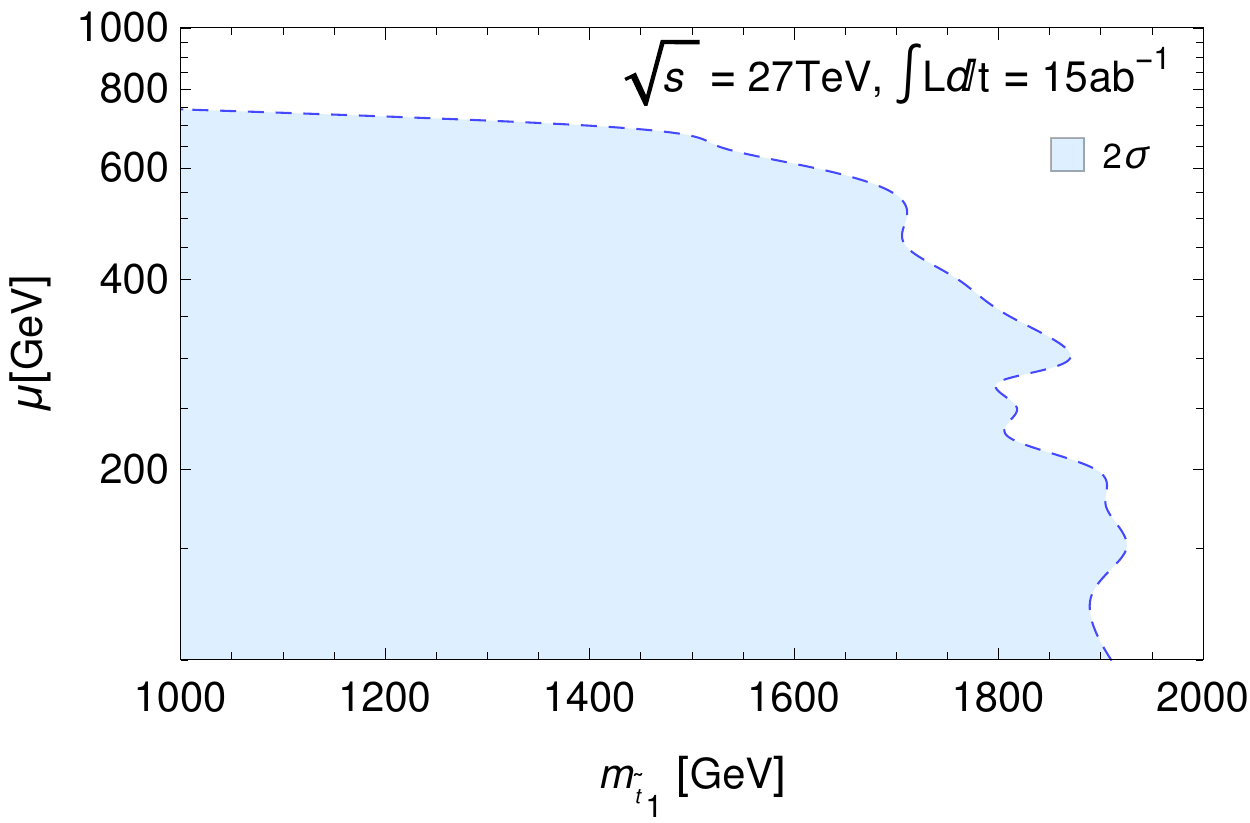}}
		\centerline{(a)}
	\end{minipage}
	\hfill
	\begin{minipage}{0.48\linewidth}
		\centerline{\includegraphics[width=8cm,height=8cm]{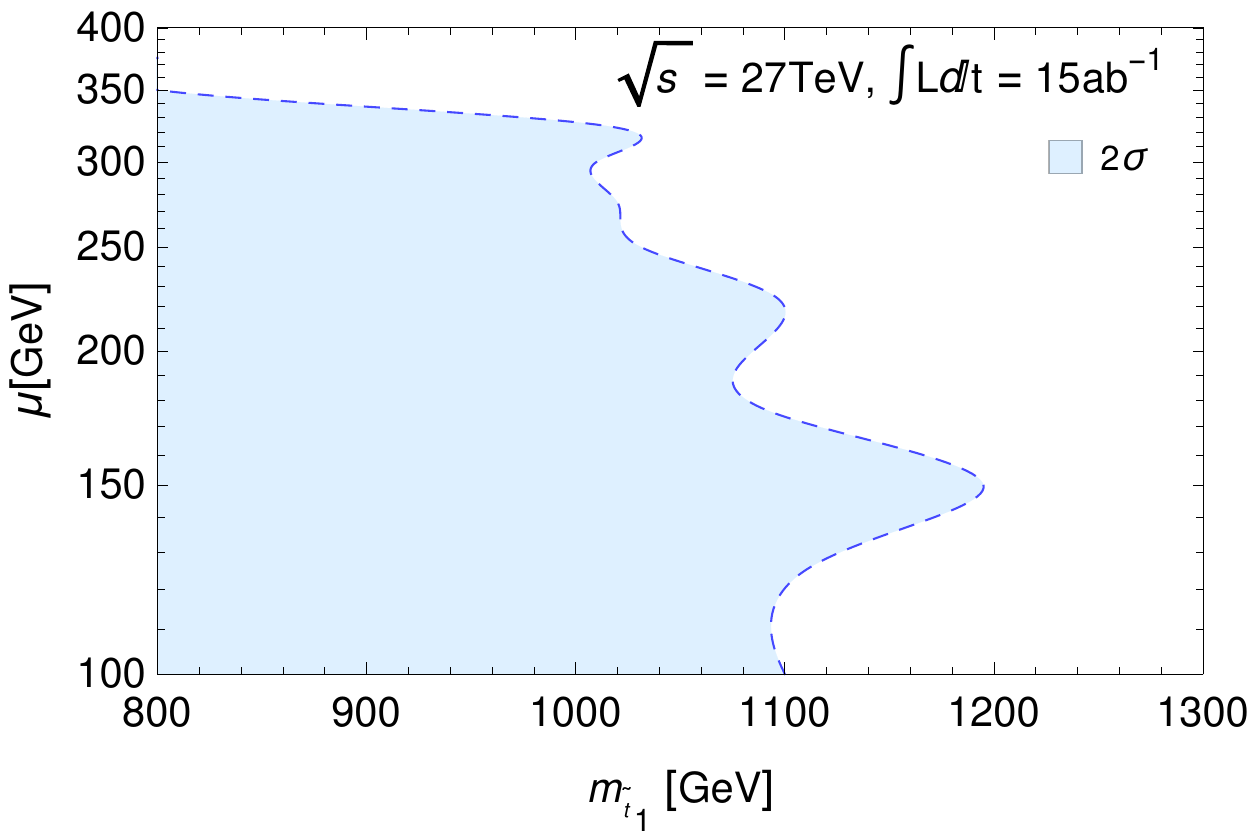}}
		\centerline{(b)}
	\end{minipage}
	\caption{The statistical significance $S/\sqrt{B}$ for the processes: (a) leptonic channel and (b) hadronic channel, on the plane of stop mass $m_{\tilde{t}_{1}}$ versus the higgsino mass parameter $\mu$ at the HE-LHC.}
	\label{fig:contour12}
\end{figure}

The same formula $\alpha=S/\sqrt{B}$ is adopted to evaluate the statistical significance and the contour on the plane of $\mu$ versus $m_{\tilde{t}_1}$ is shown in FIG.~\ref{fig:contour12}\,(b) to present the $2\sigma$ exclusion limits for the hadronic channel. The higgsino mass parameter and stop mass can be excluded at $2\sigma$ to $m_{\tilde{t}_1}<1200$ GeV and $\mu<350$ GeV, respectively. It should be noted due to the large background for the hadronic channel, the sensitivity of this channel is worse than that of the leptonic channel, as can be seen from the contours in FIG.~\ref{fig:contour12}.

In both cases of leptonic and hadronic channels, the systematic uncertainties caused by high pile-up will lead to lower statistical significance and we can resort to real performance of upgraded detectors. And new developed analysis methods including the machine-learning ones~\cite{Albertsson:2018maf,Abdughani:2019wuv,Ren:2017ymm,Caron:2016hib} may be used to improve our results in such searches at hadron colliders.

\section{CONCLUSION}

In this work, we investigate the sensitivity of the leptonic and hadronic channel from single stop production in a simplified MSSM scenario at the 27 TeV LHC with the integrated luminosity of ${\cal L} = 15~\text{ab}^{-1}$. As a complementary study to the traditional pair production, the single stop production presents some unique signatures from which we construct several useful kinematic variables to distinguish the signal from backgrounds. $2\sigma$ exclusion limits for both cases are presented: $m_{\tilde{t}_1}<1900$ GeV and $\mu<750$ GeV for the leptonic channel, while $m_{\tilde{t}_1}<1200$ GeV and $\mu<350$ GeV for the hadronic channel.

\section*{Acknowledgement}
Xu-Xu Yang would like to thank Yuchao Gu for helpful discussion. This work was supported by the National Natural Science Foundation of China (NNSFC) under grant No. 11705093, and by the Jiangsu Planned Projects for Postdoctoral Research Funds, Grant No.2019K197.

\end{document}